%% file: paper.tex
\documentclass[runningheads]{svmult}

\usepackage{mathptmx}

\usepackage{makeidx}   
\usepackage{graphicx}  
\usepackage{subeqnar}  
\usepackage{multicol}  
\usepackage{physprbb}  
\makeindex             

\newcommand{\msun}{\mbox{$\mathcal{M}_\odot$}}
\newcommand{\mbthree}{\mbox{$M_B$(3)$_0$}}
\newcommand{\mgal}{\mbox{$M$(gal)$_0$}}
\newcommand{\halpha}{\mbox{H$\alpha$}}
\newcommand{\hbeta}{\mbox{H$\beta$}}
\newcommand{\hgamma}{\mbox{H$\gamma$}}
\newcommand{\hdelta}{\mbox{H$\delta$}}
\newcommand{\wlr}{\mbox{\sc wlr}}
\newcommand{\teff}{$T_{\!\mbox{\scriptsize \em eff}}$}
\newcommand{\teffq}{$T_{\!\mbox{\scriptsize \em eff}}^4$}
\newcommand{\fglr}{\mbox{\sc fglr}}
\newcommand{\ph}{\phantom{0}}

\begin{document}
\title*{Blue Supergiants as a Tool for Extragalactic Distances --
Empirical Diagnostics\thanks{Invited review at the International
Workshop on {\em Stellar Candles for the Extragalactic Distance
Scale}, held in Concepci\'{o}n, Chile, December 9--11, 2002.  To be
published in: {\em Stellar Candles}, Lecture Notes in Physics
(http://link.springer.de/series/lnpp), Copyright: Springer-Verlag,
Berlin-Heidelberg-New York, 2003 }}

\toctitle{
Blue Supergiants as a Tool for Extragalactic Distances}
%
%
\titlerunning{Blue Supergiants as a Tool for Extragalactic Distances}
%
\author{Fabio Bresolin}
\authorrunning{Fabio Bresolin}
%
%
\institute{Institute for Astronomy, University of Hawaii, 
Honolulu HI 96822, USA\\ 
{\scriptsize\em bresolin@ifa.hawaii.edu}}

\maketitle              

\begin{abstract}
Blue supergiant stars can be exceptionally bright objects in the
optical, making them prime targets for the determination of
extragalactic distances.  I describe how their photometric and
spectroscopic properties can be calibrated to provide a measurement of
their luminosity. I first review two well-known techniques, the
luminosity of the brightest blue supergiants and, with the aid of
recent spectroscopic data, the equivalent width of the Balmer
lines. Next I discuss some recent developments concerning the
luminosity dependence of the wind momentum and of the flux-weighted
gravity, which can provide, if properly calibrated, powerful
diagnostics for the determination of the distance to the parent
galaxies.
\end{abstract}


\section{Introduction}

Massive stars can reach, during certain phases of their post-main
sequence evolution, exceptional visual luminosities, approaching
$M_V\sim-10$ in extreme cases. It is thus natural to try and use them
as standard candles for extragalactic studies, as was realized long
ago by Hubble.  For this contribution I will concentrate on the blue
supergiants\index{blue supergiants}, a rather broad but useful
definition for the stars contained in the upper part of the H--R
diagram and with spectral types O, B and A.  This includes `normal'
supergiants (Ia) and hypergiants (Ia$^+$), as well as more exotic
objects such as the Luminous Blue Variables (LBV's)\index{LBV}.  Very
bright stars can also be found among the yellow
hypergiants\index{yellow supergiants}, however their identification in
extragalactic systems is more problematic, because their intermediate
color coincides with that of numerous Galactic foreground dwarfs.

\begin{figure}[ht]
\begin{center}
\includegraphics[width=.8\textwidth]{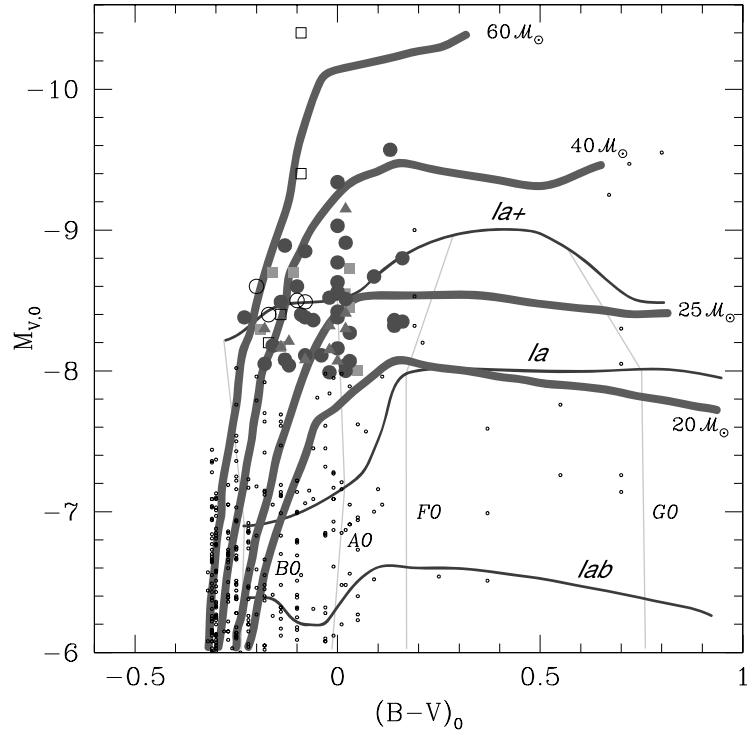}
\end{center}
\caption[]{In this color-magnitude diagram the brightest stars having
$M_V<-8$ in the Milky Way ({\it squares}), LMC ({\it circles}) and SMC
({\it triangles}) are shown with the larger symbols. Open symbols
refer to confirmed or candidate LBV's with magnitude variations
smaller than 0.5\,mag. The small points represent Galactic stars
fainter than $M_V=-8$ and/or of spectral type F and later. Schematic
evolutionary models at solar metallicity and various ZAMS masses from
\cite{meynet94} are shown by the thick lines. The grid giving
luminosity classes and spectral types is from \cite{lang1992}}
\label{cm}
\end{figure}

From the point of view of the extragalactic distance scale, it is not
the most massive, intrinsically most {\em luminous} O-type stars
($M_{bol}\leq-11$) which are appealing. Because of the decrease of the
bolometric correction with temperature from O to A stars down to
$\sim7000$\,K, the {\em visually brightest} supergiants found in
galaxies are mostly \mbox{25--40\,\msun}\/ mid-B to early-A type
stars, with $M_{bol}$ between $-8$ and $-9$ \cite{massey98}. This can
be seen from Table\,\ref{stars}, which is a (probably incomplete)
compilation of the visually brightest blue stars in the Milky
Way\index{Milky Way}, LMC\index{LMC} and SMC\index{SMC}, excluding in
general LBV's with maximum light amplitudes larger than 0.5\,mag. The
label {\sc lbv} in the last column identifies known or suspected
LBV's, from the compilation of \cite{vangenderen01}.  Sources for the
photometry and spectral types are given as a footnote to the table,
however in a few cases the data have been updated with more recent
determinations. The absolute visual magnitudes have generally been
corrected for extinction by assuming the spectral type vs. $B-V$ color
index relation in the MK system given by \cite{lang1992} and
$A_V=3.1\times E_{B-V}$.  Stars brighter than $M_V=-8.0$ in all three
galaxies are plotted in the color-magnitude (c-m)
diagram\index{color-magnitude diagram} of Fig.\,\ref{cm}, together
with the loci of Ia$^+$, Ia and Iab stars, and stellar tracks for 60,
40, 25 and 20\,\msun\/ from \cite{meynet94}.

\begin{table}
\caption{The visually brightest blue stars in the Milky Way, LMC and SMC}
\begin{center}
\renewcommand{\arraystretch}{1.}
\setlength\tabcolsep{15pt}
\scriptsize
\begin{tabular}{c c c c}
\hline\noalign{\smallskip}
\multicolumn{2}{c}{Star ID} & $M_{V,0}$ & Spectral Type \\
\noalign{\smallskip}
\hline

\noalign{\smallskip}
\multicolumn{4}{l}{\bf Milky Way \hfill $M_{V,0}\leq-8.0$ \hfill \phantom{$m-M=18.5$}}\\
\noalign{\smallskip}
\multicolumn{1}{c}{HD} & \multicolumn{1}{c}{other} \\
\cline{1-2}
\noalign{\smallskip}
\input{milkyway.table}

\noalign{\medskip}
\hline

\noalign{\smallskip}
\multicolumn{4}{l}{\bf LMC \hfill $M_{V,0}\leq-8.5$ \hfill $m-M=18.5$}\\
\noalign{\smallskip}
\multicolumn{1}{c}{HD} & \multicolumn{1}{c}{Sk} \\
\cline{1-2}
\noalign{\smallskip}
\input{lmc.table}

\noalign{\smallskip}
\hline

\noalign{\smallskip}
\multicolumn{4}{l}{\bf SMC \hfill $M_{V,0}\leq-8.0$ \hfill $m-M=19.0$}\\
\noalign{\smallskip}
\multicolumn{1}{c}{AV} & \multicolumn{1}{c}{Sk} \\
\cline{1-2}
\noalign{\smallskip}
\input{smc.table}
\noalign{\smallskip}
\hline
\end{tabular}
\end{center}
\begin{scriptsize}
{\sc sources ---} {\sc milky way -} \cite{humphreyscat}. Distance to
HD~92207, HD~92693, HD~92964 and HD223385: \cite{garmany92}.  Cyg
OB2-12: \cite{massey91}.  LBV data from \cite{vangenderen01}.  {\sc
lmc -} \cite{rousseau78}.  {\sc smc -} \cite{av82}, \cite{lennon97}.

\end{scriptsize}
\label{stars}
\end{table}

Extragalactic stellar astronomy has quickly evolved from the
identification, via photometry and qualitative spectroscopy, of
individual bright stars mostly within galaxies of the Local Group, to
the quantitative analysis of stellar spectra well beyond the
boundaries of the Local Group.  The observation and analysis of
extragalactic supergiants has important ramifications for the study of
massive stellar evolution with mass loss, supernova progenitors,
stellar instabilities near the upper boundary of the stellar
luminosity distribution, and chemical abundances. Here I review four
different techniques concerning the use of blue supergiants in the
context of measuring extragalactic distances. Two of them have quite a
long history:

\begin{itemize}
\item{\em luminosity of the brightest blue supergiants}

\item{\em equivalent width of the Balmer lines}
\end{itemize}

\noindent
while the two remaining ones are based on more recent developments in
the analysis of stellar winds and the atmospheres of blue supergiant
stars:

\begin{itemize}
\item{\em the wind momentum--luminosity relationship}

\item{\em the flux-weighted gravity--luminosity relationship}

\end{itemize}

\section{The Luminosity of the Brightest Blue Supergiants as \\a
Standard Candle}\index{brightest blue stars}

Since the pioneering work of Hubble \cite{hubble36} a great amount of
efforts have been devoted to the calibration of the luminosity of the
visually brightest blue\index{blue supergiants} and red supergiant
stars\index{red supergiants} in nearby galaxies as a distance
indicator. This work culminated in a series of papers by A.~Sandage,
R.~Humphreys and others in the 1970--80's on the bright stellar
content of galaxies in the Local Group and in a handful of more
distant late-type spirals (see reviews by \cite{sandage86},
\cite{humphreys88} and, more recently, \cite{rozanski94}).  While the
brightest red supergiants were soon recognized as a more accurate
secondary standard, thanks to the smaller dependence of their
brightness on the parent galaxy luminosity, here I will briefly
summarize the work concerning the brightest blue stars, generally of
types from late B to A. Note that a photometric color selection
criterion $(B-V)_0<0.4$ isolates supergiants of spectral type earlier
than F5.  Even if nowadays this method is not considered sufficiently
accurate when compared with the best available extragalactic distance
indicators, it has been adopted also during the past decade whenever
observational material on more accurate distance indicators (Cepheids,
TRGB, SN Ia, etc.) was lacking.

Some of the main difficulties in using the luminosity of the brightest
stars in galaxies as a standard candle were recognized by Hubble
himself, namely the unavoidable confusion between real `isolated'
stars and unresolved small stellar clusters or H\,{\sc ii} regions,
and the presence of foreground objects in the Galaxy. While the latter
problem is easily solved by avoiding stars of intermediate color in
the c-m diagram, the former is much more subtle, eventually becoming
the main criticism to the bright blue star method raised by Humphreys
and collaborators \cite{humphreysaaronson87},
\cite{humphreysaaronson87b}, who, with stellar spectroscopy in some
of the nearest galaxies, revealed the composite nature of many of
those objects which were previously considered to be the brightest
stars.

Hubble's original calibration of the mean absolute magnitude of the
three brightest stars in a galaxy, \mbthree, introduced as a more
robust measure of the visually most luminous stars than the single
brightest star, was flawed (he adopted $<M_{pg}>\,\simeq-6.3$, about 3
magnitudes too faint), which was partly responsible for the large
value he found for the expansion rate of the universe.

The dependence of \mbthree\/ on the parent galaxy luminosity
[\mbthree\/ $\propto$ \mgal], an effect already discussed by Hubble
and by Holmberg, was first investigated in detail by \cite{sandage74}
as part of a series of papers on the brightest stars in resolved
spiral and irregular galaxies, with distances calibrated via
observations of Cepheids (see \cite{sandage96}, and references
therein). The existence of such a correlation hampers the use of the
luminosity of the brightest blue stars as a standard candle. Moreover,
the standard deviation of a single observation as measured by
\cite{sandage74} was $\simeq0.5$\,mag, much larger than their quoted
0.1\,mag for the standard deviation of the mean of the three brightest
red supergiants. The latter were later also found to obey a dependence
on the parent galaxy luminosity, albeit with a shallower slope.

The \mbthree--\mgal\/ relationship has been customarily interpreted as
a statistical effect, since more luminous and larger galaxies can
populate the stellar luminosity function up to brighter magnitudes
than smaller galaxies. A flattening of this relation might be detected
in galaxies brighter than \mgal\/ $=-19$ \cite{sandage86},
corresponding to the total luminosity of large spirals, in which the
observed limit is simply imposed by the luminosity of the brightest
post-main sequence B- and A-type stars in the H--R diagram. Therefore,
while large, late-type spiral galaxies, such as M101, may contain
stars as bright as $M_B\simeq-10$, the brightest blue stars in dwarfs
like NGC~6822 or IC~1613 are found at $M_B\simeq-7$.  Numerical
simulations by \cite{schild83} and \cite{greggio86} have provided
support for the statistical interpretation, making variations in the
stellar luminosity and mass function among galaxies unnecessary to
explain the observed trend.

\begin{figure}[ht]
\begin{center}
\includegraphics[width=.9\textwidth]{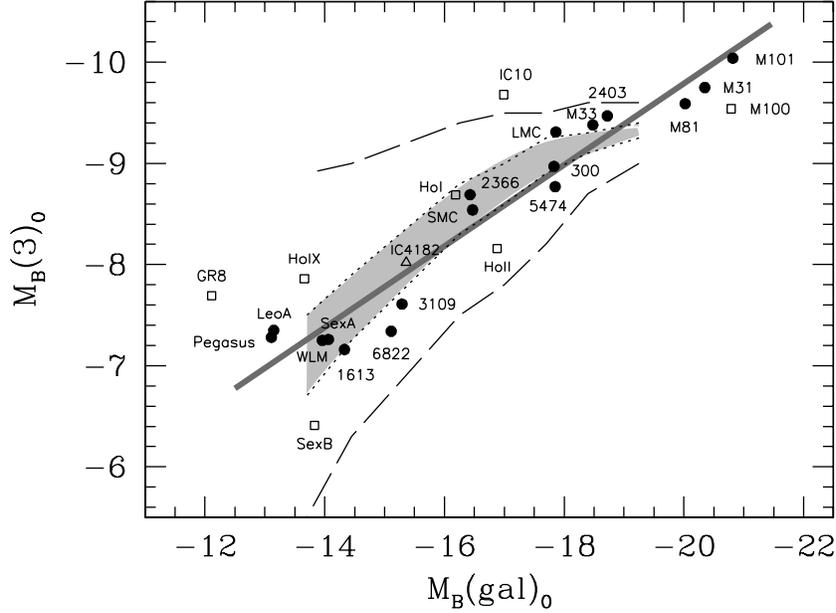}
\end{center}
\caption[]{The relationship between the average magnitude of the three
brightest blue stars and the magnitude of the parent galaxy. Data from
\cite{georgiev97}, with minor updates on the distances. Full dots
refer to the calibration galaxies, while the open symbols are used for
the additional test galaxies for which a Cepheid distance is
available. The straight line represents the linear regression to the
calibration points. The numerical simulations showing the 50\,\% and
99.5\,\% limits of the probability distribution ({\it dotted} and {\it
long-dashed} lines, respectively) are from \cite{schild83}}
\label{blue3}
\end{figure}

Among the most recent compilations of the brightest blue stars in
nearby resolved galaxies is that of \cite{georgiev97}, based on
updated stellar photometry of galaxies included in previous works by
\cite{piotto92}, \cite{karachentsev94} and \cite{rozanski94}. The
resulting relation between \mbthree\/ and parent galaxy total
luminosity is shown in Fig.\,\ref{blue3}, where different symbols are
used for 17 {\em standard} galaxies and a few {\em test} galaxies
(only those with available Cepheid distances from the list of
\cite{georgiev97} are shown, together with IC~4182
\cite{sandage96}). In this plot, distances for some of the galaxies
have been updated from the results of the HST Key Project, as
summarized by \cite{freedman01} (as an aside, no systematic study of
the brightest stars in the whole sample analyzed by the Key Project
has been published).  The standard galaxies define a linear
regression:

\begin{equation}
M_B(3)_0 = -1.76 \;(\pm 0.45) + [0.40 \;(\pm 0.03)] \;M_B(\rm{gal})_0
\end{equation}

with a standard deviation $\sigma(M_B)=0.26$. The rather small
dispersion is, at least partly, a result of the particular selection
of the `standard' galaxies made by \cite{georgiev97}.  In fact,
$\sigma(M_B)\simeq0.6$ is obtained from the data of \cite{piotto92}
and \cite{rozanski94}.  Differences in the treatment of foreground and
internal extinction exist between different authors. Furthermore,
\cite{sandage86} has advocated the use of the irregular blue variables
among the brightest blue stars, a view strongly opposed by
\cite{humphreysaaronson87}.  We must also note that consensus still
has to be reached concerning the choice of the individual brightest
blue stars in the most luminous galaxies in Fig.\,\ref{blue3}. For
example, spectroscopy of bright objects in M81 by \cite{zickgraf96}
has revealed that none of the seven brightest supergiant candidates
could be confirmed as a single star, imposing a fainter upper limit
for
\mbthree. The points in Fig.\,\ref{blue3} corresponding to the other
luminous galaxies, M31 and M101, are likely to be affected by similar
problems, and could therefore also be revised to lower \mbthree\/
values.

Numerical simulations such as those by \cite{schild83}, shown by the
dotted (50\,\% limits of the probability distribution) and long-dashed
(99.5\,\%) curves, can explain both the trend, which however is
predicted not to be linear, and the dispersion in the
observational data, at least up to the maximum galaxy brightness
considered in the models. It appears that removing objects with
Cepheid distances from the linear regression as done by
\cite{georgiev97} (those shown here by the open symbols) might not be fully
justified, since a considerable dispersion at the low-luminosity end
is expected from the incomplete filling of the stellar luminosity
function. However, considerations on the evolutionary status of some
of the dwarf galaxies might provide some justification for the removal
of some of the data points.

The general conclusion we can draw from these results is that distance
moduli to individual galaxies cannot be determined from the simple
photometry of bright blue stars to better than at least 0.5\,mag
(0.9\,mag according to \cite{rozanski94}). This is larger than
$\sigma(M_B)$, as a result of the strong dependence of \mbthree\/ on
\mgal.  Moreover, the necessity of spectroscopic confirmation of
the brightest stars must be stressed. Outside of the Local Group,
after the initial efforts at moderate resolution in M81, NGC~2403,
M101 \cite{humphreysaaronson87b}, \cite{zickgraf96}, high-quality
spectroscopy of the bright stellar content of galaxies has come within
reach of modern equipment on 8m-class telescopes at increasing
distances. As an example, I cite the work by \cite{bresolin01} and
\cite{bresolin02} in NGC~3621 and NGC~300, which will be discussed
later in this paper.

To conclude the section on extragalactic distances based on the
luminosity of the brightest blue stars, a handful of works published
in the last decade can be highlighted:

\noindent
\emph{-- brightest stars in galaxies with radial velocities
$<500$\,km\,s$^{-1}$:} a project aiming at the measurement of the
distance of a large number of (mostly dwarf) resolved galaxies in the
Local Volume ($v_{rad}<500$\,km\,s$^{-1}$) has been carried out since
1994 by Karachentsev and collaborators, using a \mbthree\/ calibration
obtained by \cite{karachentsev94} (see \cite{sharina99},
\cite{drozdovsky00} and references therein). Recently the TRGB method
is being used \cite{kara02}.

\noindent
\emph{-- brightest stars in Virgo galaxies:} brightest star candidates
have been detected from ground-based images taken under excellent
seeing conditions in two Virgo spirals, NGC~4523 \cite{shanks92} and
NGC~4571 \cite{pierce92}. Distances of 13 ($\pm2$) and 14.9 ($\pm1.5$)
Mpc were derived, respectively, from yellow and blue supergiants. The
brightest stars in a third galaxy in Virgo, M100, were discussed by
\cite{freedman94}, based on HST WFPC2 images. The Cepheid distance
from the HST Key Project is 14.3 ($\pm0.5$) Mpc.

\noindent
\emph{-- additional galaxies in the field (D\,$\sim$\,7-8 Mpc):} the
brightest blue and red supergiants have been used by \cite{sohn98},
\cite{sohn96a} and \cite{sohn96b} to measure distances to a few spiral
galaxies, including NGC~925 and NGC~628, adopting the calibration of
\cite{rozanski94}.


\section{Spectroscopic Diagnostics: Equivalent Width of \\the
Balmer Lines}\label{ew_chapter}

The spectroscopic approach\index{blue supergiants!spectroscopy}
alleviates the major difficulties of the photometric method described
in the previous section. Small clusters, close companions and H\,{\sc
ii} regions can be easily identified from line profiles, composite
appearance of the spectrum and presence of nebular lines. In addition,
the analysis of the spectral diagnostics (equivalent widths, line
profiles, continuum fluxes) can provide detailed information on
element abundances, spectral energy distributions, wind outflows and
stellar reddening.

The discovery of a relationship between stellar optical spectral lines
and luminosity dates back to the 1920's, when the character of the
lines, diffuse vs. sharp, and their strength were found to correlate
with the absolute magnitude of stars of type A and B \cite{adams22},
\cite{adams23}, \cite{edwards27}, \cite{williams29}. The hydrogen
Balmer lines in particular were soon recognized to play an important
role in connection with the problem of measuring stellar luminosities
using spectra, a fact which continues to hold true even for the most
recent techniques involving the spectral analysis of blue supergiants.
The luminosity effect on the width of the Balmer lines\index{Balmer
lines!luminosity effects} derives from their dependence on the
pressure (Stark broadening), as realized by
\cite{hulburt24} and \cite{struve29}, with a line absorption
coefficient in the wings proportional to the electron pressure (and
also dependent on the temperature).  As a result, narrower and weaker
lines are formed with decreasing pressure and surface gravity, and
consequently with increasing luminosity.

I will not discuss here additional, somewhat related methods,
including: $i$) the relationship between $M_V$ and the strength of the
O~{\sc i} triplet at $\lambda\sim7774$\,\AA, which holds for the A--G
spectral types \cite{arellano02}; $ii$) the strength and the
effective wavelength of the Balmer jump, as in the Barbier, Chalonge
\& Divan classification system \cite{chalonge73}, and $iii$)
photometric indexes centred on selected Balmer lines, such as the
$\beta$ index used by \cite{crawford78}, \cite{crawford79} and
\cite{zhang83}. Another luminosity diagnostic for B9--A2
supergiants, the strength of the Si~{\sc ii}~$\lambda\lambda6347,6371$
lines, was proposed by \cite{rosendhal74}, but \cite{feast76} showed
that this indicator breaks down for bright SMC stars, as a likely
effect of the reduced metallicity.

Work by \cite{petrie65} on the equivalent width of the \hgamma\/ line,
W(\hgamma)\index{Balmer lines!\hgamma}\index{equivalent
width!\hgamma}, led to a calibration of its relationship with absolute
magnitude, lower values of W(\hgamma) being found for high-luminosity
stars. A spectral type dependence among the B and A stars of different
luminosity classes was also detected. The cut-off at the bright end of
this early calibration ($M_V>-7$) was imposed by the scarcity of
supergiant stars with known distances.

Refinements to the calibration of this technique were introduced by
\cite{balona74} and by \cite{hutchings66}. The latter used a
W(\hgamma)--$M_V$ calibration based on Galactic stars to estimate the
distance to the Magellanic Clouds, thus pioneering stellar
spectroscopy as a way to determine extragalactic distances (see also
\cite{crampton79}).  More recent calibrations have been proposed by
\cite{millward85} (O--A dwarfs and giants), \cite{walker85} and
\cite{hill86} (supergiants), accounting for the spectral type
dependence. Among the applications, I recall the work by
\cite{azzopardi87}, who used W(\hgamma) to determine luminosity
classes for a large number of stars in the SMC.

\begin{figure}[ht]
\begin{center}
\includegraphics[width=.99\textwidth]{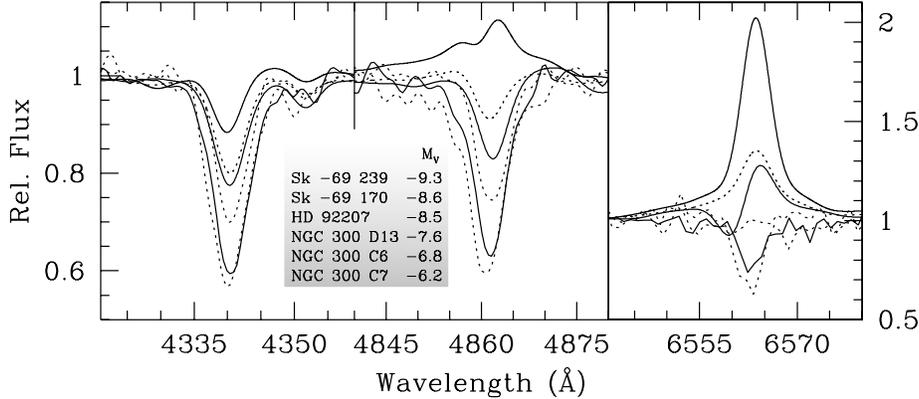}
\end{center}
\caption[]{Examples of \hgamma\/ (left), \hbeta\/ (middle) and
\halpha\/ (right) line profiles in blue supergiants of different
visual brightness (decreasing from top to bottom, as indicated in the
legend) in the LMC, Milky Way and NGC~300. The LMC and Galactic
spectra (courtesy N. Przybilla and R. Kudritzki) have been degraded to
the 5\,\AA\/ resolution of the NGC~300 data}
\label{balmer}
\end{figure}

Correlations between Balmer lines of blue supergiants and stellar
luminosity are not restricted to the use of \hgamma.  A strong
luminosity effect on the \halpha\/ line \index{Balmer
lines!\halpha}\index{equivalent width!\halpha}was found from
narrow-band photometry of Galactic early-type stars by
\cite{andrews68}. Later \cite{tully84} turned their attention to the
equivalent width of the \halpha\/ and \hbeta\/ lines in a sample of B--A
supergiants in the LMC and SMC. The availability in these
extragalactic systems of a large number of blue supergiants up to
extreme luminosities, all at a common distance and with small
reddening, is a major advantage for calibration purposes.  The
\halpha\/ and \hbeta\/ lines are in emission for the visually
brightest blue stars in the Clouds, as recognized since the early
stellar spectroscopic work in these galaxies \cite{feast60}, a
signature of the presence of extended atmospheres and mass
loss\index{mass loss}
through stellar winds. The luminosity effect is particularly strong in
\halpha, which in late-B and early-A supergiants begins to show a
clear emission nature, mostly with a characteristic P-Cygni profile,
around $M_V=-7$ \cite{rosendhal73}. The filling of the line profiles
by stellar wind becomes progressively smaller as one proceeds to
Balmer lines of higher order, so that \hgamma, \hdelta\/, etc. are
increasingly better diagnostics of stellar surface gravity. Examples
of \halpha, \hbeta\/ and \hgamma\/ line profiles are shown in
Fig.\,\ref{balmer} for stars of different visual brightness, from
\mbox{$M_V=-9.3$} to $-6.2$ in the LMC (the two brightest objects),
the Milky Way (HD~92207) and NGC~300 (the three fainter objects), all
plotted at the same intermediate spectral resolution.  Excellent
examples of higher resolution profiles of Balmer lines of B--A
supergiants can be found in the papers by \cite{rosendhal73},
\cite{kud99} and \cite{verdugo99}.

For extragalactic distance studies it is essential that the scatter in
the relationships between observables be small. In the case of the
equivalent width of \halpha\/ and \hbeta\/ vs. magnitude, the rms
scatter found by \cite{tully84} for about 40 B5--A0 supergiants in the
LMC was 0.3--0.4\,mag. However, the scatter doubles for similar stars in
the SMC, an effect attributed to a metallicity dependence of the mass
loss rate. When \hgamma\/ and \hdelta\/ are considered
\cite{tully86} a $\sim0.5$\,mag dispersion is found in the LMC. On
the other hand, \cite{walker85} and \cite{hill86}, using their
W(\hgamma)--$M_V$ calibration, claim a probable error $\simeq0.2$\,mag
(standard deviation $\simeq0.3$\,mag), for a single observation.
However, we note that in the latter two works stars brighter than
$M_V=-8$ are excluded from the calibration, somewhat reducing its
usefulness for extragalactic work.

Since the mid-1980's not much work has been published about new
applications of the W(\hgamma)--$M_V$ relationship, possibly because of
the somewhat uncertain results obtained in the Clouds and, most of
all, because of the lack of high-quality spectra for blue supergiants
in galaxies beyond the Magellanic Clouds.  With the availability of
8--10m telescopes in recent times the spectroscopy of a large number of
stars well beyond the Local Group boundaries has become feasible, and
new calibrations and tests of spectroscopic luminosity diagnostics are
likely to appear in future years. Several projects are underway within
our group and others to use the current generation of multi-object
spectrographs (FLAMES, FORS and VIMOS at the VLT, DEIMOS at Keck, GMOS
at Gemini) to observe a large fraction of the bright stellar content
in galaxies of the Local Group and beyond.

\begin{figure}[ht]
\begin{center}
\includegraphics[width=.9\textwidth]{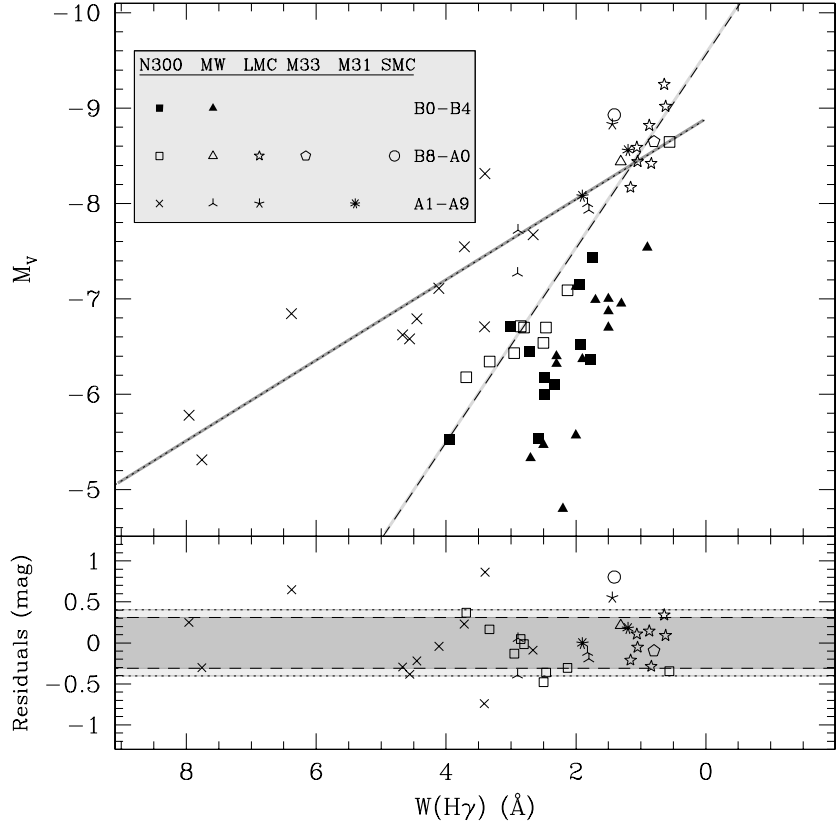}
\end{center}
\caption[]{(Top) The W(\hgamma)--$M_V$ relationship for B--A supergiants
in NGC~300, Milky Way, Magellanic Clouds, M31 and M33. The sample has
been divided into three classes: B0--B4 ({\it full symbols}), B8--A0
({\it open}) and A1--A9 ({\it crosses}). The regression lines for the
latter two are shown.  (Bottom) Residual plot from the regressions for
B8--A0 ({\it open symbols}) and A1--A9 supergiants ({\it
crossed}). The standard deviation is shown by dashed and dotted lines,
respectively}
\label{hgamma}
\end{figure}

A first, modern version of the W(\hgamma)--$M_V$ relation for B and A
supergiants, based on CCD spectra collected within our group, is
reproduced in Fig.\,\ref{hgamma}.  The sample shown contains objects
from the following galaxies: NGC~300 \cite{bresolin02}, Milky Way
\cite{kud99}, LMC and SMC \cite{przybilla}, M33 and M31
\cite{mccarthy95}, \cite{mccarthy97}. It has been subdivided,
somewhat arbitrarily, into three separate classes according to the
stellar spectral type range: early B (B0--B4, full symbols), B8--A0
(open symbols) and A1--A9 (crosses).

The slope of the empirical relationship in Fig.\,\ref{hgamma} becomes
shallower for the later spectral types, as indicated by the
regressions corresponding to the B8--A0 (dashed line) and A1--A9
(dotted line) classes. This trend with spectral type is well-known
from previous work, with a maximum W(\hgamma) at a given $M_V$ around
type F0 for supergiants \cite{hutchings66}. Contrary to the
calibration by \cite{hill86}, the new diagram is populated up to very
bright magnitudes, $M_V\simeq-9$. The relationship defined by the
B8--A0 subgroup is rather tight, with a standard deviation of about
0.3\,mag (bottom panel of Fig.\,\ref{hgamma}), and is given by

\begin{equation}
M_V = -9.56 \;(\pm0.15) + [1.01 \;(\pm0.07)] \;W(\hgamma)\;.
\end{equation}

The scatter for the later A-type supergiants is 50\,\% larger. A
further subdivision of this broad class might reveal tighter
correlations, but currently this is prevented by the small number of
objects available. We note the rather large discrepancy of the only A0
Ia supergiant plotted for the SMC (AV 475) from the regression line,
with an \hgamma\/ line too strong for its magnitude.  A metallicity
effect cannot be excluded at this stage to explain this discrepancy.
The measurements of W(\hgamma) in SMC supergiants by
\cite{azzopardi87}, combined with $M_V$'s obtained from magnitudes and
spectral types in the catalog by \cite{av82}, are in general agreement
with those shown in Fig.\,\ref{hgamma}, although they show a larger
scatter. This might be, at least partly, related to the necessity of
redefining the spectral type classification at low metallicity, as
shown by \cite{lennon97}.

To conclude, by restricting the analysis to a narrow range in spectral
types (B8 to A0) the scatter about the mean W(\hgamma)--$M_V$ relation
is on the order of 0.3\,mag, which makes this spectroscopic technique
rather appealing for its simplicity and accuracy, at least for
metallicities comparable to that of the LMC and larger, whenever
moderate resolution spectra of a sufficient number of B8--A0
supergiants in a given galaxy are available. Additional tests at lower
metallicity (for example in the SMC) should be carried out to verify
the dependence on chemical abundance.


\section{The Wind Momentum--Luminosity Relationship}\label{wlrsec}

\begin{figure}[ht]
\begin{center}
\includegraphics[width=.9\textwidth]{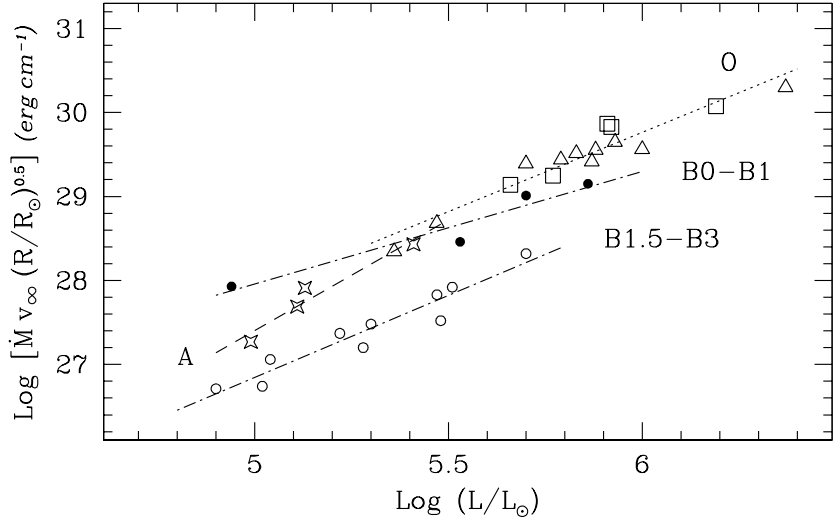}
\end{center}
\caption[]{Spectral type dependence of the \wlr\/ for Galactic
supergiant stars. Different symbols are used for the different
spectral type ranges, and regression lines are drawn. Adapted from
\cite{kud99}, with data from the same paper. For O stars the blanketed
model results of \cite{puls2002} and
\cite{herrero2002} have been used}
\label{wlr_all}
\end{figure}

The discovery and empirical verification of a relationship between the
intensity of the stellar wind momentum and the luminosity of massive
stars is certainly one of the foremost successes of the theory of line
driven winds, which is presented in R. Kudritzki's contribution in
this volume (see also \cite{kud2000} for a
review). The predicted Wind Momentum--Luminosity Relationship (\wlr)
can be written as:

\begin{equation}
\log D_{mom} = \log D_0 + x \log\frac{L}{L_\odot}
\end{equation}

where the linear regression coefficients $D_0$ and $x$ are derived
empirically from observations of O, B and A stars at known distances.
The modified wind momentum $D_{mom}=\dot{M} v_\infty
(R/R_\odot)^{0.5}$, i.e. the product of the mass-loss rate, wind
terminal velocity and square root of stellar radius, is determined
spectroscopically, once the magnitude of the star is measured. A
spectral type dependence is found, as shown in Fig.\,\ref{wlr_all}, as
a result of the different ionic species driving the wind at different
stellar temperatures.  In fact the slope $x$ corresponds to the
reciprocal of the exponent $\alpha^\prime$ of the power-law describing
the line-strength distribution function.  The predicted value lies
around $\alpha^\prime=1/x=0.6$.  The effects of metallicity $Z$ are
also to be empirically verified, while the predictions for the
dependence of both $\dot{M}$ and $v_\infty$ indicate that
approximately $D_{mom}
\propto Z^{0.8}$ \cite{vink2001}.

Since all massive and luminous blue stars show signs of mass loss, it
is interesting to take advantage of this through the \wlr\/ as a way
to measure extragalactic distances.  In practice, one needs to obtain
the mass-loss rate from the \halpha\/ line fitting, and the photospheric
parameters (temperature, gravity and chemical composition) from
optical absorption lines in the blue optical spectral region
(4000--5000\,\AA). An empirically calibrated \wlr\/ as a function of
metallicity would then allow the determination of distances, once the
apparent magnitude, reddening and extinction are known. The latter
quantities can be derived from the observed spectral energy
distribution and theoretical models calculated with the appropriate
photospheric parameters. The method is backed by a strong theoretical
framework, especially for the hot O stars, which allows us to deal
quantitatively with the spectral diagnostics of blue supergiants.
However, its observational application requires careful spectroscopic
analysis and modeling, which can make it intimidating at first. On the
other hand, the same analysis provides us with a large amount of
information on the physics of massive stars.  Here I will briefly
summarize the results obtained so far concerning the calibration of
the \wlr, by discussing the O stars separately from the B--A
supergiants, referring the reader to the papers cited above for the
theoretical aspects and for details on how the stellar and wind
parameters are extracted from the observational data.

\subsection{O Stars}
\begin{figure}[ht]
\begin{center}
\includegraphics[width=.9\textwidth]{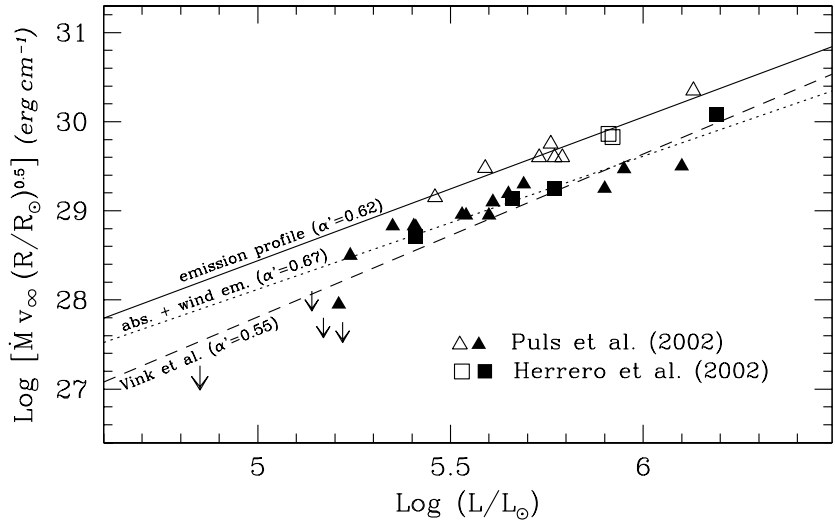}
\end{center}
\caption[]{The \wlr\/ for Galactic O stars, with data from
\cite{puls2002} and \cite{herrero2002}. The O star sample has been
subdivided according to the character of the \halpha\/ profile, either
in emission ({\it open symbols}) or in absorption partly filled by
wind emission ({\it full symbols}). The linear fits to the two
sub-samples are shown, together with the theoretical models by
\cite{vink00}. The slope of the line strength distribution
function $\alpha^\prime=1/x$ is indicated for each regression}
\label{wlr_o}
\end{figure}

For O stars the reference work remains the paper by Puls et
al. \cite{puls96}, in which theoretical results from unified model
atmosphere calculations were used to measure mass-loss rates from
\halpha\/ line profiles. Their method overcomes the inaccuracies
related to the use of the equivalent width of the \halpha\/ line from
the wind emission, corrected for photospheric contribution, as done by
\cite{leitherer88} and \cite{lamers93}. For a sample of Galactic and
Magellanic Clouds stars the relevant parameters were measured from
spectra in the UV ($v_\infty$, based on the P-Cygni profiles of strong
resonance lines) and in the optical (\teff, $\log g$), and from the
analysis of the
\halpha\/ line profile ($\dot{M}$, wind velocity law). Separate
relations were found by Puls and collaborators for different
luminosity classes, the supergiants having larger wind momenta than
giants and dwarfs at a given luminosity. Moreover, reduced wind
momenta were measured in the Magellanic Clouds when compared with the
Milky Way stars, as a manifestation of a metallicity effect on the
strength of the wind. More recently, \cite{herrero2002} have
determined the \wlr\/ for O stars in a single Galactic association,
Cyg OB2, thus reducing the uncertainty in stellar distances as a
source of scatter in the calibration. Recent improvements in the
stellar models allowed these authors to account for the effects of
line blocking and blanketing from metals. As explained by
\cite{puls2002} the consequent cooler temperature scale for O stars,
when compared with \teff\/ calibrations based on unblanketed
atmospheres, modifies the \wlr\/ significantly, leading in particular
to an indication that the luminosity class dependence might be no
longer present, in agreement with the theoretical predictions by
\cite{vink00}, but instead that wind clumping might mimick higher
mass-loss rates in the more extreme cases, affecting preferentially
stars with an \halpha\/ profile in emission.  Fig.\,\ref{wlr_o}
illustrates these results, where the Galactic sample of
\cite{puls2002} and the Cyg OB2 stars of \cite{herrero2002} have been
divided according to the nature of the \halpha\/ line profile, i.e. in
emission (open symbols) or in absorption partly filled by wind
emission (full symbols). The resulting tight relations are clearly
displaced from one another by about 0.3 dex, with the less extreme
objects (\halpha\/ in absorption, optically thin winds) lying very
close to the theoretical relationship defined by \cite{vink00}. The
hypothesis of clumping to explain the apparently higher mass-loss
rates in stars with \halpha\/ in emission is very appealing, but
requires further observational confirmation, with the UV-to-IR
spectral analysis of a large number of O stars.

\subsection{B and A Supergiants}
Even if the \wlr\/ calibration for the O stars is probably the best
available to date, and the one upon which most observational and
theoretical work has been concentrated, it is the late-B and the A
supergiants which offer the largest potential as extragalactic
distance indicators. In fact, these stars are in general much less
affected by crowding and confusion problems, which afflict O stars,
normally found in OB associations and/or within H\,{\sc ii} region
complexes.  Besides, they attain brighter magnitudes in the optical
($M_V\simeq-9$ for A hypergiants, vs. $M_V\simeq-7$ for the brightest
O Ia stars), as the result of a combination of stellar evolution
across the upper H--R diagram at roughly constant luminosity and
smaller bolometric corrections in the covered temperature range
(\teff~$\simeq13000-7500$\,K). Their extreme luminosities make their
analysis less affected by the presence of fainter companions, which
can be, if bright enough to be of importance, detected by their
spectral signatures. As a downside, a large fraction of the brightest
supergiants are photometrically variables at the 0.1--0.2\,mag level
\cite{vangenderen89}, \cite{grieve86}, and this should be taken into
account when trying to use them as distance indicators.

With the modern spectroscopic capabilities at 8m-class telescopes,
A-type supergiants should be within reach of a quantitative analysis
out to distance moduli $(m-M)\simeq$\,30--31, i.e. out to the distance
of the Virgo Cluster.  A great amount of work, however, still needs to
be done regarding the identification of suitable targets on one hand,
and the calibration of the \wlr\/ on the other.

\subsubsection*{Selecting the Candidates}

For those resolved galaxies where extensive stellar classification
work has not already been carried out, which in practice, with a few
exceptions, means all galaxies outside of the Local Group, as well as
a considerable fraction of Local Group members, spectral types of blue
supergiant candidates must be found from scratch, overall a rather
lengthy process. This is accomplished by first obtaining stellar
photometry in two or more bands, typically $B$ and $V$, from CCD
images, and subsequently concentrating the spectroscopic follow-up on
those isolated objects in the color and magnitude range expected for
blue supergiants. We have found it very helpful, if not necessary, to
obtain also narrow-band
\halpha\/ images, to limit as much as possible the contamination on
the final stellar spectra from nebular lines. This is particularly
true in the late-type, star-forming galaxies which are the natural
targets for the application of the
\wlr. The availability of large-format CCD mosaics at several 2--4
meter telescopes around the globe, covering as much as half a degree
or more on the sky in a single exposure, is making the photometric
surveys required for target selection time-efficient even when
considering a large number of nearby galaxies.

\begin{figure}[ht]
\begin{center}
\includegraphics[width=.99\textwidth]{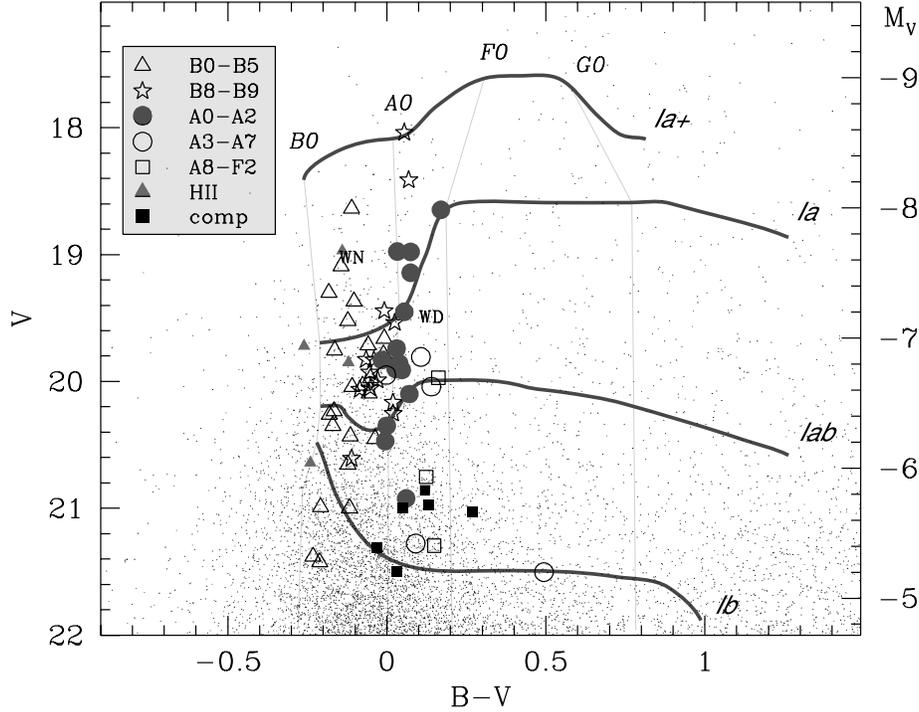}
\end{center}
\caption[]{Color-magnitude diagram of NGC 300. The supergiants studied
spectroscopically by \cite{bresolin02} have been divided into separate
ranges of spectral type, according to the legend in the upper
left. Observed H\,{\sc ii} regions and objects with composite spectra
are also indicated, together with the locations of a foreground white
dwarf (WD) and a WN11 star (WN). The magnitude-color calibration for
Ia$^+$, Ia, Iab and Ib stars is the same as in Fig.~\ref{cm}}
\label{N300cm}
\end{figure}

The photometric selection is always affected at some level by the
presence of `intruders' in the c-m diagram (unresolved stellar groups,
unidentified small H\,{\sc ii} regions, stars with greatly different
extinction, and more exotic objects), and can be verified only {\em a
posteriori}, once the spectroscopy has been carried out.  This is
becoming less of a problem, with the availability of multiobject
spectrographs, like FLAMES at the VLT, which allow the simultaneous
observation of hundreds of spectra.

As an example of some recent results, Fig.\,\ref{N300cm} shows the c-m
diagram of the Sculptor Group galaxy NGC~300, obtained from 2.2m/WFI
CCD images at La Silla by W.~Gieren, where the objects we have studied
spectroscopically with FORS1 at the VLT have been indicated
\cite{bresolin02}. With our original selection criterion
($-0.3<B-V<0.3$, $V<21.5$) we intended to isolate late-B and early-A
supergiants. The confirmed stellar types, from B0 to F2, correspond
rather well with the expected location in the c-m diagram. The few
pre-selected H\,{\sc ii} regions have rather blue $B-V$ colors, while
some objects characterized by a composite spectrum are found at faint
magnitudes. More interesting interlopers are a foreground Galactic
white dwarf and a WN11 emission line star, the latter analyzed by
\cite{bresolinWN}. The high success rate in the case of NGC~300 has
been made possible by the careful analysis of the broad-band and
\halpha\/ images, in combination with the modest distance (2\,Mpc),
and likely by the small foreground+internal reddening, even though a
number of bright supergiants were found to be contaminated by nebular
emission in subsequent
\halpha\/ spectra used for the mass-loss determination.  Work similar
to the one described for NGC~300 has been carried out at the VLT by
our group in two additional galaxies, NGC~3621 \cite{bresolin01} and a
second galaxy in Sculptor, NGC~7793 (work in preparation). We are also
securing wide-field images of about a dozen nearby ($D<7$\,Mpc)
galaxies from La Silla and Mauna Kea, as well as obtaining HST/ACS
imaging of selected fields in NGC~300, NGC~3621 and M101 for accurate
stellar photometry of confirmed and candidate blue supergiants.

\subsubsection*{Calibration of the WLR}

The current calibration of the \wlr\/ for Galactic A-type supergiants,
given by \cite{kud99}, rests on only four stars, because of the
difficulty to measure reliable distances and reddening corrections for
this type of objects when located in the Milky Way. In their paper, 14
early B supergiants were also studied, but I will not include them in
the following discussion because of their somewhat lower appeal (being
fainter) for extragalactic distances work. Apparently they also do not
conform to the same relationship as the A supergiants, as seen in
Fig.\,\ref{wlr_all}. A subset of these stars (those in the B1.5--B3
spectral type range) were found to have low wind momenta when
compared to the O and early B stars, an anomaly which is not currently
understood within the theoretical framework. I simply mention here
that several observing programs are underway on the early B
supergiants in Local Group galaxies, utilizing optical spectroscopy to
derive photospheric parameters and abundances \cite{smartt01},
\cite{trundle02} and UV spectroscopy to measure wind velocities
\cite{urbaneja02}, \cite{bresolinUV}, in order to obtain a better
understanding of their wind properties.

From the paucity of Galactic calibrators it is clear that the sample
must be enlarged by observing additional bright B--A supergiants in
nearby galaxies, before any attempt to measure independent distances
with the \wlr\/ is made.  The first high resolution spectra of A
supergiants in M33 and M31, two in each galaxy, have been obtained
with the Keck telescope by
\cite{mccarthy95} and \cite{mccarthy97}, respectively, demonstrating
that, at least at distances smaller than 1 Mpc, all stellar wind
parameters ($\dot{M}$, $v_\infty$ and the exponent $\beta$ of the wind
velocity law) can be satisfactorily obtained from fitting the
\halpha\/ line profile. The paper on the M31 stars, in particular,
shows several examples of how varying the wind parameters influences
the calculated profiles.  The wind analysis from the \halpha\/ line
must still be carried out in a larger number of supergiants in these
two galaxies, however. The M33 spiral, in particular, appears to be an
ideal target, because of its moderate distance and its favorable
inclination on the plane of the sky. The radial chemical abundance
gradient in this galaxy is such that it will allow the investigation
of the metallicity effect on the \wlr.  We are currently planning to
secure spectra at the required resolution (about 1--2\,\AA) with Keck
equipped with the new multiobject spectrograph DEIMOS.

In 1999 we have started to use the FORS spectrograph at the VLT in
order to increase the sample of extragalactic A-type supergiants
having a spectroscopic coverage. Two are the principle goals: to
determine stellar abundances, which are important for galactic and
stellar evolution investigations, and to measure wind momenta, for an
experimental verification and calibration of the \wlr.  The very first
targets, a handful of blue supergiants in NGC~6822, were observed
during the commissioning phase of FORS1, confirming that even at
relatively low resolution (5\,\AA) quantitative spectroscopy can be
successfully carried out, with regard to both stellar metallicities
and mass-loss rates \cite{muschielok99}.  An important step was taken
with the subsequent analysis of stellar spectra in NGC~3621
\cite{bresolin01}, so far the most distant galaxy, at $D\simeq6.7$\,Mpc, for
which spectroscopy of individual supergiants has been
published. Unfortunately, red spectra covering \halpha\/ are still
unavailable for this galaxy, so that the mass-loss rate of the blue
supergiants discovered could not be determined, except for a highly
luminous A1 Ia star ($M_V=-9.0$), for which the wind-affected \hbeta\/
has been used.

After having verified the potential of the available instrumentation,
we have turned to more nearby galaxies, NGC~300 and NGC~7793 in the
Sculptor Group, combining our project with an investigation of their
Cepheid content \cite{pietr02}. While the FORS2 data for NGC~7793 have
just recently come in, the analysis of the blue supergiants in NGC~300
has already provided important results concerning the classification
and the first A-type supergiant abundance estimates
\cite{bresolin02}, the A-supergiant \wlr\/ (in preparation) and the
metallicity of the B-type stars \cite{urbanejanew}.

\begin{figure}[ht]
\begin{center}
\includegraphics[width=.8\textwidth]{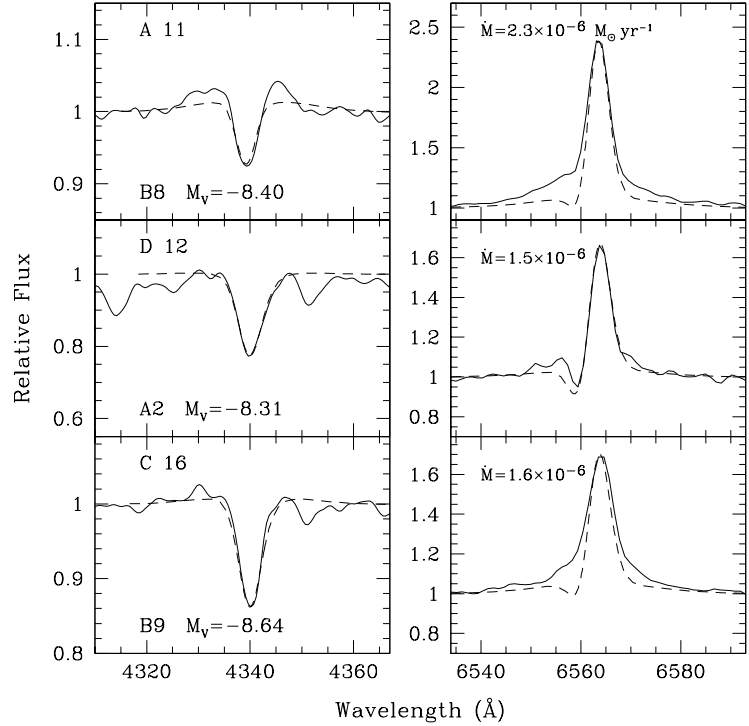}
\end{center}
\caption[]{\hgamma\/ (left) and \halpha\/ (right) line profile fits
({\it dashed lines}) to the spectra of three bright supergiants in
NGC~300, obtained with FASTWIND \cite{santolaya97}.  The
identification number from \cite{bresolin02} is given in the upper
left of each plot. Spectral types and absolute magnitudes are indicated in the
lower part of the \hgamma\/ panels, and mass-loss rates in
M$_\odot$\,yr$^{-1}$ in the \halpha\/ panels}
\label{hafits}
\end{figure}

The wind analysis has been carried out so far for six A-type
supergiants (B8 to A2) from red spectra obtained at the VLT/FORS1 in
September 2001. The unblanketed version of the non-LTE line formation
code {\sc fastwind} \cite{santolaya97} has been used to fit the
observed \halpha\/ line profiles. The major drawback when using
spectroscopic data at moderate resolution is that the terminal
velocity cannot be determined from the line fits, forcing us to adopt
a reasonable estimate for it, $v_\infty=150$\,km\,s$^{-1}$. Although
such a velocity is typical for well studied Galactic A supergiants,
this assumption is currently the major source of uncertainty in the
calculation of the wind momentum in the NGC~300 sample.
Fig.\,\ref{hafits} illustrates profile fits to
\hgamma\/ (mostly sensitive to gravity) and \halpha\/ (mass-loss rate)
for three bright A-type supergiants in NGC~300. As can be seen, in
some cases we cannot reproduce the observed extended electron
scattering wings of \halpha\/, however these features arise in the
photosphere, and do not affect the mass-loss rate determination, which
is carried out from the peak of the emission. We can also neglect the
discrepancies in the blue absorption part of the P-Cygni profiles.
This feature is often affected by variability in high luminosity
objects, but the corresponding wind momentum variations are contained
within the scatter of the \wlr\/ \cite{kud99conf}.

\begin{figure}[ht]
\begin{center}
\includegraphics[width=.9\textwidth]{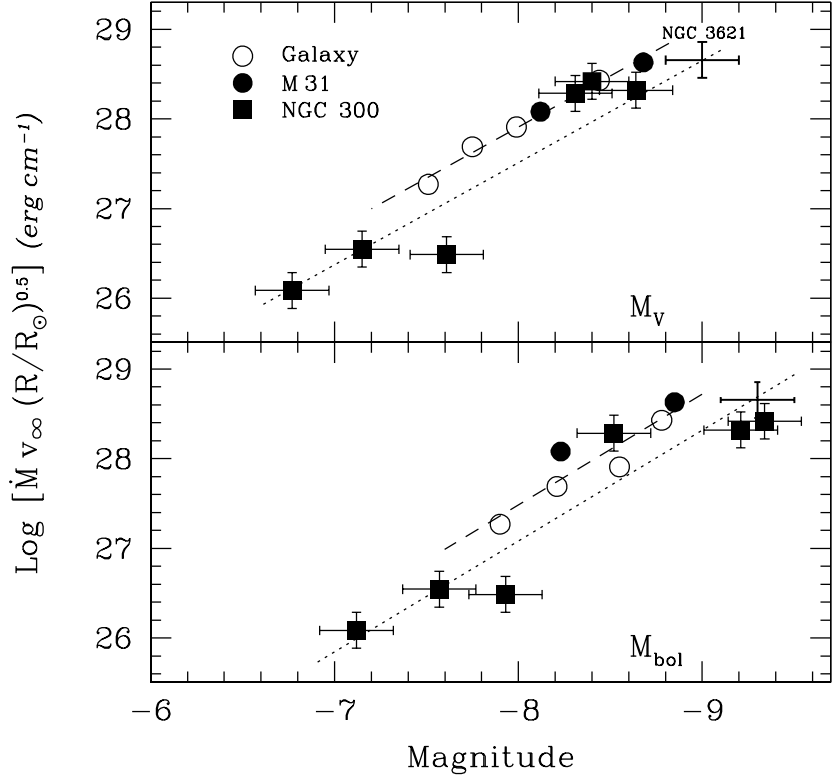}
\end{center}
\caption[]{The \wlr\/ for A-type supergiants, represented in terms of
$M_V$ (top) and $M_{bol}$ (bottom). The stellar objects are drawn from
samples of blue supergiants in the Milky Way, M31, NGC~300 and
NGC~3621. The linear fits to the Galactic and M31 supergiants are
given by the dashed lines. A theoretical scaling factor is applied to
provide the expected relations at 0.4\,$Z_\odot$ ({\it dotted lines})}
\label{wlr_a}
\end{figure}

The \wlr\/ determined for the NGC~300 stars analyzed so far, expressed
as a relation between modified wind momentum and both $M_V$ and
$M_{bol}$, is shown in Fig.\,\ref{wlr_a}. The diagrams also include
the four Galactic A supergiants studied by \cite{kud99}, the two stars
in M31 from \cite{mccarthy97} and the only star in NGC~3621 for which
we were able to determine the mass-loss rate from the \hbeta\/ line
(again, assuming here $v_\infty=150$\,km\,s$^{-1}$).

The metallicity for the NGC~300 stars, which lie all at a similar
galactocentric distance, is estimated, from the known H\,{\sc ii}
region oxygen abundance gradient \cite{zaritsky94} and from the
appearance of the stellar spectra, to be around
$0.4\,\pm\,0.1$\,$Z_\odot$. The NGC~3621 star has a comparable
metallicity.  If we adopt an empirical `calibration' of the \wlr\/ at
solar metallicity as provided by a fit to the Milky Way and M31
points, given by the dashed lines, we can then scale to the lower
metallicity using a $Z^{0.8}$ dependence. This is shown by the dotted
lines in Fig.\,\ref{wlr_a}.

Let us concentrate on the $D_{mom}-M_{bol}$ relation, since the sample
of objects considered varies in spectral type from B8 to A3, implying
different bolometric corrections by up to 0.7\,mag. The corresponding
plot in Fig.\,\ref{wlr_a} shows a rather well defined \wlr\/ for the
Galactic and M31 stars, with a scatter of about 0.2 dex in the
modified wind momentum. All the points corresponding to NGC~3621 and
NGC~300, except one, agree with the expected approximate location for
$Z=0.4\,Z_\odot$. We have yet to carry out the chemical abundance
analysis of the discrepant point, and we have currently no obvious
explanation for its location in the diagram. Despite this, it appears
that even with the moderate resolution spectra at our disposal we can
well define the \wlr\/ in a galaxy at a distance of 2\,Mpc. Of course
this result is only preliminary, in the sense that we are still
lacking a real calibration of the \wlr\/ which we could use to
determine extragalactic distances. Analyzing a larger sample of
objects, with a range in metallicity from $Z_\odot$ down to
$0.1\,Z_\odot$, still remains among our top priorities.  The obvious
candidates for such work are stars in the Magellanic Clouds, and in a
number of nearby spirals and irregulars rich in blue supergiants, such
as M33, M31 and NGC~6822. Moreover, an improved spectral analysis
which takes into account the blanketing and blocking effects of metals
in the stellar atmosphere must be carried out.

\section{A New Spectroscopic Method: the Flux-Weigthed
Gravity--Luminosity Relationship}

A very promising luminosity diagnostic for blue supergiants has been
very recently proposed by \cite{kudnew}, based on the realization that
the fundamental stellar parameters, surface gravity and effective
temperature, are predicted to be tightly coupled with the stellar
luminosity during the post-main sequence evolution of massive
stars. While quickly crossing the upper H--R diagram from the main
sequence to the red supergiant phase both the mass and the luminosity
of late-B to early-A supergiants remain roughly constant, so that by
postulating an approximate mass-luminosity relationship ($L\sim
\mathcal{M}^{3}$) we derive:

\begin{equation}
M_{bol} = a\log(g/T_{\!\mbox{\scriptsize \em eff}}^4) + b
\end{equation}

with $a\simeq-3.75$.  This equation defines a {\em Flux-weighted
Gravity--Luminosity Relationship} (\fglr), since the quantity
$g$/\teffq\/ can be interpreted as the flux-weighted gravity.  Even
for the most massive stars in the range of interest, 30--40\,\msun,
the mass-loss rate is still small enough that during the typical
timescale of this evolutionary phase its effects on the
\fglr\/ are negligible, thus justifying our assumption of mass
constancy.  This is also true, to first order, for the effects of
differing metallicities and rotational velocities, as confirmed by the
results of evolutionary models \cite{meynet00}.

The \fglr\/ is conceptually very simple, and its empirical
verification only requires, besides the visual magnitude, the
measurement of $\log g$ and \teff\/ from the stellar spectrum.  A
complication arises from the difficulty of measuring these parameters
with sufficient accuracy in extremely bright supergiants and
hypergiants. Recently, however, sophisticated non-LTE modeling of A
supergiants can provide us with tools to determine the stellar
parameters with high reliability \cite{venn01}, \cite{przybilla01},
\cite{przybilla01b}.

\begin{table}
\caption{Adopted temperature scale for supergiants}
\begin{center}
\renewcommand{\arraystretch}{1.}
\setlength\tabcolsep{10pt}
\scriptsize
\begin{tabular}{c r}
\hline\noalign{\smallskip}
Spectral Type & \teff\phantom{0} \\
\noalign{\smallskip}
\hline
\noalign{\smallskip}
B8      &       12000 \\
B9      &       10500 \\
A0      &       9500 \\
A1      &       9250 \\
A2      &       9000 \\
A3      &       8500 \\
A4      &       8350 \\
\noalign{\smallskip}
\hline
\end{tabular}
\end{center}
\label{temps}
\end{table}

As a first test, we have analyzed the spectra of blue supergiants from
a number of Local Group galaxies, observed as part of our
investigation of the \wlr, together with our FORS data on NGC~300 and
NGC~3621, as described in Sect.~\ref{wlrsec}. Effective temperatures
were estimated from the observed spectral types, according to the
correspondence shown in Table\,\ref{temps}, except for the lower
metallicity objects in SMC, M33 and NGC~6822, for which \teff\/ was
derived from the non-LTE ionization equilibrium of Mg~{\sc i/ii} and
N~{\sc i/ii}. Surface gravities were measured from a simultaneous fit
to the higher Balmer lines (\hgamma, \mbox{\hdelta, ...)}, therefore
minimizing the wind emission contamination on the lower-order hydrogen
lines. Despite the moderate spectral resolution of the FORS data,
which prevents us to fit the wings of the spectral lines, we are able
to achieve a similar {\em internal} accuracy as for the higher
resolution spectra by fitting the line cores, about 0.05 dex in $\log
g$. In practice at low resolution we are fitting the Balmer line
equivalent widths to measure stellar gravities, and the \fglr\/
technique is therefore reminiscent of the
\hgamma\/--$M_V$ relation discussed in
Sect.~\ref{ew_chapter}. However, we are now considering a larger
number of Balmer lines, and including a correction for the temperature
dependence through the \teffq\/ term. The bolometric correction and
the intrinsic color, which allow us to account for reddening and
extinction, are also determined from the spectral analysis.

\begin{figure}[ht]
\begin{center}
\includegraphics[width=.9\textwidth]{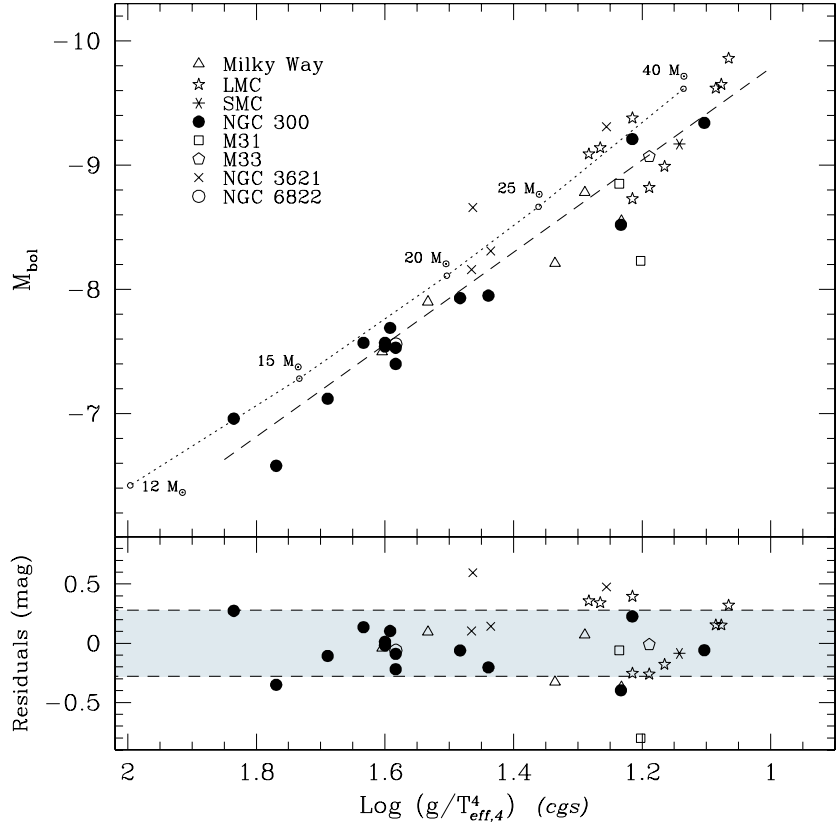}
\end{center}
\caption[]{(Top) The relationship between the flux-weighted gravity
$g$/\teffq\/ (\teff\/ in units of $10^4$\,K) and $M_{bol}$ for B8--A4
supergiants in Local Group galaxies, NGC~300 and NGC~3621. The dashed
line represents the linear regression to all the data points.  The
models at \teff\,=\,$10^4$\,K from the evolutionary calculations
accounting for rotation ($v_{in}=300$\,km\,s$^{-1}$) and at solar
metallicity of
\cite{meynet00} are connected by the dotted line, labeled with the
corresponding ZAMS masses. (Bottom) Residuals from the regression
shown in the top panel, with the standard deviation indicated by the
dashed line}
\label{fglr}
\end{figure}

The results are displayed in Fig.\,\ref{fglr}, where the dashed line is
the linear regression:

\begin{equation}
M_{bol} = 3.71 \;(\pm0.22)\, \log(g/T_{\!\mbox{\scriptsize \em
eff,4}}^4) - 13.49 \;(\pm0.31)
\end{equation}

with a standard deviation $\sigma=0.28$ (note that \teff\/ is
expressed in units of $10^4$\,K). This is a very encouraging
result for extragalactic work. As can be seen from Fig.\,\ref{fglr},
the points corresponding to the blue supergiants in NGC~300 define a
tight relationship ($\sigma=0.20$), despite the intermediate
resolution of the spectra available. For NGC~3621 the \fglr\/ with
just four stars would suggest a distance slightly larger than the
assumed Cepheid distance, however the upcoming photometry from HST/ACS
is required before we can draw any firm conclusion.

The slope of the empirical \fglr\/ reproduces very well the
theoretical expectations from the evolutionary models of
\cite{meynet00}, while the offset can be interpreted, for example, as
a systematic effect on the determination of the stellar parameters,
which would not be important in a strictly differential analysis. The
apparent increase of the dispersion at high luminosities might be a
consequence of the larger role played by stellar rotation and/or
metallicity on the flux-weighted gravity.  Still, with a standard
deviation of 0.3--0.35\,mag and with the analysis of about ten blue
supergiants in a given galaxy we could be able to determine a mean
relationship, and therefore the distance modulus, with an accuracy of
$\sim0.1$ magnitudes. Our aim is to apply the \fglr\/ method to
distances of up to $m-M=30.5$, where stars brighter than $M_V=-8$ can
be observed with the high-efficiency spectrographs currently available
at 8--10m telescopes.

Is the \fglr\/ going to replace the \wlr\/ as a more promising
distance indicator for blue supergiants?  The advantages of the
\fglr\/ are multiple. Medium resolution spectra seem to be sufficient,
and no red spectrum covering \halpha\/ is required, thus reducing the
amount of time at the telescope. All the stellar classification work,
the chemical abundance analysis and the fit to the higher Balmer lines
can be carried out in the blue spectral region, which is also less
contaminated by night sky lines.  Work on the \wlr\/, however, should
continue, especially for early-B stars, for which the assumptions upon
which the \fglr\/ rests could fail.

The results shown here are just the starting point in the study of the
\fglr. A detailed and extensive calibration work must be carried out
in Local Group galaxies, where with instrument like FLAMES (VLT) or
DEIMOS (Keck) we can obtain hundreds of blue supergiant spectra. Such
observations, besides providing us with an accurate calibration of the
\fglr, will also improve our understanding of the effects on stellar
evolution of those parameters, like metallicity, mass-loss and angular
momentum, which are relevant for a theoretical explanation of the
relationship.

As a concluding remark, one may ask the question why we should insist
in using spectroscopic methods for blue supergiants, such as the
\wlr\/ or the \fglr, as extragalactic distance indicators, when other
well-tested photometric techniques (e.g. Cepheids and Tip of the Red
Giant Branch) promise to obtain perhaps higher accuracy and to reach
more distant objects with arguably less observational efforts. The
answer is of course that a greater physical insight is gained from the
analysis of the stellar spectra, allowing us to determine {\em for
each individual stellar target} the crucial parameters of metallicity
and reddening. The discrimination against unresolved companions or
small clusters is also possible through the appearance of the spectra.
Because of the importance of each of these factors in the application
of the main distance indicators used nowadays, possibly leading
to some systematic errors in the distance scale if uncorrected for, it
is important to continue our efforts to measure a number of
`spectroscopic' distances to galaxies of the nearby universe.


\section{Acknowledgments}
I wish to thank the Organizing Committee, W. Gieren and D. Alloin in
particular, for inviting me to this workshop and for the hospitality,
and R.P.~Kudritzki and W.~Gieren for their help and encouragement.  I
am also grateful to N. Przybilla for making his spectra of Magellanic
Cloud supergiants available. This research has made use of the SIMBAD
database, operated at CDS, Strasbourg, France.


%


\end{document}

%% file: milkyway.table
		& Cyg OB2-12	& $-10.4\ph$	& B5 Ie {\sc lbv} \\
80077		&		& $-9.4\ph$	& B2/3 Ia$^+$ {\sc lbv}\\
92693		&		& $-8.73$	& A2 Ia \\
152236		& $\zeta^1$ Sco	& $-8.70$	& B1.5 Ia$^+$ \\
		& WRA977	& $-8.7\ph$	& B1.5 Ia$^+$ \\
92207		&		& $-8.55$	& A0 Ia \\
197345		& $\alpha$ Cyg	& $-8.45$	& A2 Iae \\
168607		&		& $-8.4\ph$	& B9 Ia$^+$ {\sc lbv}\\
316285		& He3-1482	& $-8.4\ph$	& BIe {\sc lbv}\\
169454		&		& $-8.29$	& B1 Ia$^+$ \\
		& MWC314	& $-8.2\ph$	& $<$B2 {\sc lbv}\\
92964		&		& $-8.17$	& B2.5 Iae \\
223385		& 6 Cas		& $-8.00$	& A3 Iae \\

%% file: lmc.table
33579	&	$-67~44$	& 	$-9.57$	&	A4 Ia$^+$ \\
269902	&	$-69~239$	& 	$-9.34$	&	B9 Iae \\
32034	&	$-67~17$	& 	$-9.03$	&	B9 Iae \\
270086	&	$-69~299$	& 	$-8.91$	&	A1 Ia$^+$ \\
269546	&	$-68~82$	& 	$-8.89$	&	B3 Iab \\
269923	&	$-69~247$	& 	$-8.85$	&	B6 Iab \\
269857	&	$-68~131$	& 	$-8.80$	&	A9 Ia \\
269781	&	$-67~201$	& 	$-8.77$	&	A0 Iae \\
269331	&	$-69~93$	& 	$-8.67$	&	A5 Ia \\
268654	&	$-69~7$		& 	$-8.63$	&	B9 Iae \\
268835	&	$-69~46$	& 	$-8.60$	&	B8p \\
269128	&	$-68~63$	&	$-8.6\ph$ &	B2.5 Ia$^+$ {\sc lbv}\\
269661	&	$-69~170$	& 	$-8.56$	&	B9 Ia$^+$ \\
268718	&	$-69~16$	& 	$-8.52$	&	B9 Iabe \\
268946	&	$-66~58$	& 	$-8.51$	&	A0 Ia \\
37836	&	$-69~201$	&	$-8.5\ph$ &	B pec {\sc lbv}\\

%% file: smc.table
 475	&	152	&  $-9.15$	& A0 Ia$^+$ \\
 136	&	54	&  $-8.41$	& A0 Ia \\
        &	56	&  $-8.32$	& B8 Ia$^+$ \\
 315	&	106	&  $-8.30$	& A0 Ia \\
 78	&	40	&  $-8.30$	& B1.5 Ia$^+$\\
 443	&	137	&  $-8.21$	& B2.5 Ia \\
 56	&	31	&  $-8.17$	& B2.5 Ia \\
 65	&	33	&  $-8.15$	& B8 Ia$^+$ \\
 48	&	27	&  $-8.08$	& B5 Ia \\
 76	&	39	&  $-8.07$	& B9 Ia$^+$ \\